# Emergence of superconductivity on the border of antiferromagnetic order in RbMn$_6$Bi$_5$ under high pressure: A new family of Mn-based superconductors

P. T. Yang[1,2=], Q. X. Dong[1,2=], P. F. Shan[1,2=], Z. Y. Liu[1,2=], J. P. Sun[1,2], Z. L. Dun[1,2], Y. Uwatoko[3], G. F. Chen[1,2*], B. S. Wang[1,2*], J.-G. Cheng[1,2*]

[1]*Beijing National Laboratory for Condensed Matter Physics and Institute of Physics, Chinese Academy of Sciences, Beijing 100190, China*

[2]*School of Physical Sciences, University of Chinese Academy of Sciences, Beijing 100190, China*

[3]*Institute for Solid State Physics, University of Tokyo, Kashiwa, Chiba 277-8581, Japan*

= These authors contributed equally to this work.

Correspondence should be addressed to B.S.W. (bswang@iphy.ac.cn), G.F.C. (gfchen@iphy.ac.cn) or J.G.C. (jgcheng@iphy.ac.cn)

## Abstract

The advances in the field of unconventional superconductivity are largely driven by the discovery of novel superconducting systems. Here we report on the discovery of superconductivity on the border of antiferromagnetic order in a quasi-one-dimensional RbMn$_6$Bi$_5$ via measurements of resistivity and magnetic susceptibility under high pressures. With increasing pressure, its antiferromagnetic transition with $T_N \approx 83$ K at ambient pressure is first enhanced moderately and then suppressed completely at a critical pressure of $P_c \approx 13$ GPa, around which bulk superconductivity emerges and exhibits a dome-like $T_c(P)$ with a maximal $T_c^{onset} \approx 9.5$ K at about 15 GPa. Its temperature-pressure phase diagram resembles those of many magnetism-mediated superconducting systems. In addition, the superconducting state around $P_c$ is characterized by a large upper critical field $\mu_0 H_{c2}(0)$ exceeding the Pauli limit, elaborating a possible unconventional paring mechanism. The present study, together with our recent work on KMn$_6$Bi$_5$ ($T_c^{max} \approx 9.3$ K), makes $A$Mn$_6$Bi$_5$ ($A$= Alkali metal) a new family of Mn-based superconductors with relatively high $T_c$.

# Introduction

The advances in the field of unconventional superconductivity are largely driven by the discovery of novel superconducting systems as exemplified as the cuprates and iron-pnictide high-temperature superconductors[1-3]. The first-row (3d) transition-metal-based compounds are at the forefront of materials' discovery for unconventional superconductivity due to the presence of strong electronic correlations and the proximity to magnetic instability. In 2015, some of us discovered pressure-induced superconductivity with $T_c \approx 1$ K in MnP by suppressing its helimagnetic order under high pressure[4]. This finding breaks the general wisdom about Mn's antagonistic to superconductivity and has thus promoted the quest for more Mn-based superconductors with higher $T_c$. After years of explorations, we recently found that the quasi-one-dimensional (Q1D) $KMn_6Bi_5$ becomes superconducting with a relatively high $T_c^{onset}$ up to 9.3 K when its antiferromagnetic order is suppressed by pressure[5]. This finding makes $KMn_6Bi_5$ the first *ternary* Mn-based superconductor with an optimal $T_c$ about an order higher than that of MnP. Importantly, this result demonstrates that the $T_c$ of Mn-based superconductors has a potential to go higher. In addition, the obtained temperature-pressure phase diagram of $KMn_6Bi_5$, featured by a superconducting dome on the border of antiferromagnetically ordered state, resembles those of many unconventional superconducting systems associated with magnetism-mediated pairing mechanism[6-8]. Since $KMn_6Bi_5$ is one member of the Q1D $AMn_6Bi_5$ ($A$ = Alkali metal) system[9], it is natural to ask whether they consist of a new family of Mn-based superconductors. To address this issue, here we turn our attention to $RbMn_6Bi_5$.

As shown in Fig. 1(a), $RbMn_6Bi_5$ also adopts the monoclinic structure (space group $C2/m$) featured by the infinite $[Mn_6Bi_5]^-$ columns, which are composed of an outer nanotube of Bi atoms, an inner Mn-Mn bonded metallic pentagon core and a one-dimensional Mn-Mn atomic chain in the center along the $b$ axis. The counter-cation $Rb^+$ ions fill the space between the $[Mn_6Bi_5]^-$ columns, acting as the structural frame and carriers' source[9,10]. Replacing $K^+$ with a larger $Rb^+$ in $AMn_6Bi_5$ leads to anisotropic expansions of lattice parameters, i.e. the $a$ and $c$ are expanded by 1.3% and 1.9%, respectively, while the $b$ is only increased by 0.2%. This means that the $[Mn_6Bi_5]^-$ columns remain almost intact by the replacement of larger $A$-site cation while the intercolumn interactions are weakened, leading to a stronger anisotropy in $RbMn_6Bi_5$. However, at ambient pressure $RbMn_6Bi_5$ undergoes an antiferromagnetic transition at $T_N \approx 82$ K, slightly higher that of $KMn_6Bi_5$ ($T_N \approx 75$ K). Around $T_N$, the resistivity $\rho_{//}$ along the rod (i.e. $b$-axis) shows a weak kink but $\rho_\perp$ perpendicular to the rod exhibits a pronounced upturn before returning to the metallic state. First-principles calculations indicate that the density of states at Fermi level are mainly contributed from the Mn-3d electron bands in the Mn10 pentagonal antiprisms[10] and the nonlinear helical antiferromagnetic structures are found to be stable in energy and to be similar with MnP and $A_2Cr_3As_3$[11-14]. In this regard, the itinerant magnetic nature of RbMn$_6$Bi$_5$ with enhanced anisotropy and relatively low $T_N \approx 82$ K makes it a new candidate for unconventional superconductivity near the magnetic quantum critical point (QCP).

Through detailed measurements of resistivity and magnetic susceptibility at high pressures, here we find that RbMn$_6$Bi$_5$ also becomes superconducting when its antiferromagnetic order is suppressed by pressure at a critical pressure of $P_c \approx 13$ GPa. The maximal $T_c^{onset} \approx 9.5$ K is achieved at ~15 GPa. In addition, the superconducting state around $P_c$ is characterized by a large upper critical field $\mu_0 H_{c2}(0)$ exceeding the Pauli limit, elaborating a possible unconventional paring mechanism. In combination with our recent work on KMn$_6$Bi$_5$[5], the present study demonstrates that $A$Mn$_6$Bi$_5$ ($A$= Alkali metal) represent a new family of ternary Mn-based superconductors with relatively high $T_c$, and should be subjected to more experimental and theoretical investigations.

**Results**

**Physical properties at ambient pressure (AP).** The RbMn$_6$Bi$_5$ crystals are black in color with metallic luster and have a needle shape with the longest dimension along the $b$-axis. They were first characterized at AP via measurements of electrical resistivity $\rho(T)$ and magnetic susceptibility $\chi(T)$ along the rod direction. As shown in Fig. 1(b), the $\rho(T)$ of RbMn$_6$Bi$_5$ shows a metallic behavior in the whole temperature range and exhibits a clear hump anomaly around the antiferromagnetic transition at $T_N \approx 83$ K, where a sudden drop in $\chi(T)$ appears. The estimated residual resistivity ratio ($RRR \equiv \rho$ (300 K)/$\rho$ (2 K)) for the RbMn$_6$Bi$_5$ crystal is about 17, and the curie-Weiss fitting to the paramagnetic susceptibility above 100 K yields an effective moment of 1.68 $\mu_B$/Mn. These values are close to those reported in literature.[10] It should be noted that the observed $\rho(T)$ is different from that reported in Ref. 10 along the $b$-axis. The observed upturn feature indicates the inclusion of contribution from $\rho_\perp$. As shown in the inset of Fig. 1(b), a close inspection of the RbMn$_6$Bi$_5$ crystals reveals that they are composed of a bundle of thin fibers, consistent with the Q1D character of the crystal structure. Such a character makes it easy to pick up the signal perpendicular to the $b$-axis when measuring resistivity with the standard four-probe configuration on the bulk crystal along the longest dimension. Nonetheless, these above characterizations confirm the metallic nature of RbMn$_6$Bi$_5$ with an antiferromagnetic transition at $T_N \approx 83$ K.

**High-pressure resistivity (CAC).** Figure 2(a) shows the resistance $R(T)$ data of RbMn$_6$Bi$_5$ (#1) measured along the $b$-axis under various hydrostatic pressures up to 14.5 GPa in a cubic anvil cell (CAC) apparatus. The overall evolutions of $R(T)$ under high pressure are similar to those observed in KMn$_6$Bi$_5$.[5] At 2 GPa, the $R(T)$ shows a saturation behavior at high temperatures and a weak kink anomaly around $T_N$, followed by a broad hump at low temperatures. Here, $T_N$ can be determined from the peak of the

d$R$/d$T$, Fig. 2(c). With increasing pressure, the resistance is reduced progressively and the characteristic features around $T_N$ and below are weakened gradually; the $T_N$ values determined d$R$/d$T$ are enhanced slightly from 83 K at AP to ~ 93 K at 6 GPa. Upon further increasing pressure above 6 GPa, the $R(T)$ behaves differently, and the downward kink-like feature around $T_N$ evolves into a weak upward anomaly; the corresponding anomaly in d$R$/d$T$ changes from a peak at lower pressures to a broad dip at higher pressures as seen in Figs. 2(c) and S1. Similar crossover is also found in the resistance of KMn$_6$Bi$_5$, and might be associated with the modification of the antiferromagnetic structure under pressure. In this pressure range, $T_N$ is reduced gradually with pressure, reaching ~ 46 K at 12 GPa. No clear anomaly can be discerned in the $R(T)$ of 12.5 GPa, indicating that the long-range antiferromagnetic order in RbMn$_6$Bi$_5$ has been suppressed completely.

Accompanying the collapse of antiferromagnetic order in RbMn$_6$Bi$_5$, we observed a sudden drop of resistance at low temperatures, signaling the possible occurrence of superconductivity. As shown in Fig. 2(b), the $R(T)$ at 13 GPa starts to decrease at $T_c^{onset}$ ≈ 6.6 K, and zero resistance can be reached below $T_c^{zero}$ ≈ 2.3 K at 13.5 GPa. With increasing pressure, the superconducting transition moves to higher temperatures gradually; at 14.5 GPa, $T_c^{onset}$ and $T_c^{zero}$ reach about 9.2 K and 4.3 K, respectively. The observed optimal $T_c$ value of RbMn$_6$Bi$_5$ is close to that of KMn$_6$Bi$_5$ (9.3 K), but is much higher than that of MnP (~ 1 K).

**The upper critical field.** To characterize the superconducting state, $R(T)$ data under different magnetic fields were recorded at 13.5, 14, and 14.5 GPa for RbMn$_6$Bi$_5$. The $R(T)$ data at 14.5 GPa are shown in Fig. 3(a), and all the data are given in Fig. S2. With increasing magnetic field, the superconducting transition is shifted to lower temperatures and broadened up gradually owing to the magnetic-breaking effect and the flux creep effect in the vortex state. Here, we employed the criteria of 50% $R_n$ to determine $T_c$ and plotted the temperature dependence of $\mu_0 H_{c2}(T)$ in Fig. 3(b). The zero-temperature upper critical field $\mu_0 H_{c2}(0)$ is obtained by fitting the experimental data to the Ginzburg-Landau (GL) equation, $\mu_0 H_{c2}(T) = \mu_0 H_{c2}(0)[1-(T/T_c)^2]/[1+(T/T_c)^2]$[15]; then, the coherent length $\xi(0)$ can be calculated according to the relation: $\mu_0 H_{c2}(0) = \Phi_0/2\pi\xi(0)^2$, where $\Phi_0 = h_c/2e$ is the magnetic flux quantum. We find that $\mu_0 H_{c2}(0)$ increases dramatically from 9.7 T at 13.5 GPa to 18.0 T at 14.5 GPa, exceeding the Pauli limit of $\mu_0 H_p = 1.84\ T_c = 11.1$ T,[16] while the $\xi(0)$ decreases from 58.2 to 47.5 Å. The presences of large $\mu_0 H_{c2}(0)$ and small coherent length are the common features of unconventional superconductivity. We note that the $\mu_0 H_{c2}(T)$ curve displays a tail near 0 T and cannot be described well by the simple GL fitting, which should be ascribed to the multi-band effect.

**AC magnetic susceptibility under high pressure (CAC).** Figure 3(c) shows the ac magnetic susceptibility data $\chi'(T)$ for RbMn$_6$Bi$_5$ (#2) and a piece of Pb, which serves as a superconducting reference to estimate the superconducting shielding volume fraction of RbMn$_6$Bi$_5$. The pressure values in Fig. 3(c) were estimated based on the $T_c$ of Pb. As can be seen, only a sharp superconducting transition of Pb is observed at $P \leq$ 12.3 GPa, which further elaborates an excellent hydrostatic pressure environment in CAC. At $P \geq$ 13.6 GPa, in addition to the sudden drop of $\chi'(T)$ for Pb, we observed a gradual reduction of $\chi'(T)$ below $T_c^\chi$ associated with the superconducting transition of RbMn$_6$Bi$_5$. At 14.5 GPa, $T_c^\chi$ reaches about 8 K, consistent with the resistance data shown in Fig. 2(b), and the superconducting shielding volume fraction at 1.5 K increases to nearly 100 %, confirming the bulk nature of the observed superconductivity in RbMn$_6$Bi$_5$.

**High-pressure resistivity (DAC).** To track the evolution of superconducting transition at higher pressures, we also measured the $R(T)$ of RbMn$_6$Bi$_5$ (#3) by using a diamond anvil cell (DAC) up to 23.3 GPa. As seen in Fig. 3(d), the superconducting transition can be clearly detected as a drop of resistance below $T_c^{onset}$, but the zero resistance cannot be achieved in the investigated pressure range, presumably due to the non-hydrostatic pressure conditions in DAC. With increasing pressure, $T_c^{onset}$ increases monotonously from ~ 8.2 K at 13.8 GPa to a maximum of ~ 9.5 K at 15.5 GPa, and then decreases gradually to ~4.1 K at 23.3 GPa, leading to a superconducting dome as shown below.

**Temperature-pressure phase diagram.** Based on the above experimental results obtained in CAC and DAC, we can construct a temperature-pressure phase diagram for RbMn$_6$Bi$_5$ as shown in Fig. 4. With increasing pressure, $T_N(P)$ first increases moderately untill 6 GPa, where it experiences a sudden drop and then decreases gradually before it vanishes abruptly at a critical pressure of $P_c \approx$ 13 GPa. Bulk superconductivity emerges at about 12.5 GPa and zero-resistance state is achieved at 13.5 GPa. $T_c(P)$ increases monotonously to a maximal value of $T_c^{onset}$ = 9.5 K at 15.5 GPa and then reduces with further increasing pressures. These features define a pressure-induced superconducting dome on the border of long-range antiferromagnetically ordered state, Fig. 4. Such a T-P phase diagram of RbMn$_6$Bi$_5$ is very similar to that of KMn$_6$Bi$_5$ and resembles those of many unconventional superconducting systems associated with the magnetism-medicated pairing mechanism[6-8].

**Discussions.** The discovery of pressure-induced superconductivity with a relatively high $T_c$ ~ 9 K in the Q1D (K,Rb)Mn$_6$Bi$_5$ is quite encouraging and should stimulate more studies in these new Mn-based superconductors. Firstly, they represent a new class of ternary Mn-based superconductors with unique Q1D crystal structure and possible novel superconducting gene. As mentioned above, the Mn-based superconductors are quite rare owing to the strong magnetic pair-breaking effect. So far, pressure-induced superconductivity has been observed only in the binary MnP and MnSe.[4,17] In these cases, other tuning methods expect applying pressure are less effective in regulating the magnetism and inducing superconductivity. In contrast, the ternary $A$Mn$_6$Bi$_5$ with a unique Q1D structure provides more possibilities to tune the physical properties, e.g., via carrier doping or chemical substitutions at the $A$- and/or Bi-sites.

Secondly, they offer a new materials' platform to study the interplay between exotic magnetism and superconductivity. The magnetic structure of $A$Mn$_6$Bi$_5$ has not been determined experimentally. According to the density-functional-theory calculations, RbMn$_6$Bi$_5$ could adopt a complex helical antiferromagnetic structure at ambient pressure[10]. As pointed out previously, the emergence of superconductivity in a helical magnet is rare and how the spin fluctuations associated with helical order play any role in the mechanism driving superconductivity deserves in-depth studies. In this regard, it is highly desirable to determine the magnetic structure of $A$Mn$_6$Bi$_5$ at AP and then to reveal its evolution under high pressure. In addition, the information about the spin dynamics near the magnetic QCP around $P_c$ are also important to understand the superconducting mechanism.

Thirdly, considering the observed large $\mu_0H_{c2}(0)$ exceeding the Pauli limit, it is also interesting to investigate the superconducting pairing symmetry. For example, spin-triplet superconducting state has been proposed and verified by the nuclear magnetic resonance experiments in the Q1D superconductor $A_2$Cr$_3$As$_3$ with similar crystal structure[18]. Last but not least, although the Mn-based compounds are commonly believed to be antagonistic to superconductivity and thus should have a low $T_c$, the present work demonstrates that the transition temperature of Mn-based superconductors can be raised to the level of 10 K or higher. More investigations on $A$Mn$_6$Bi$_5$ and other complex Mn-based compounds are desirable to further raise the $T_c$ of Mn-based superconductors.

In summary, we report the discovery of superconductivity on the border of antiferromagnetic order in the Q1D RbMn$_6$Bi$_5$ under pressure. Bulk superconductivity emerges and display a superconducting dome with the maximal $T_c^{onset} \approx 9.5$ K at about 15 GPa. The superconducting state near the optimal $T_c$ is characterized by a large upper critical field $\mu_0H_{c2}(0)$ exceeding the Pauli limit, implying an unconventional superconducting paring mechanism. In combination with our recent work on KMn$_6$Bi$_5$, the present study demonstrates that $A$Mn$_6$Bi$_5$ ($A$= Alkali metal) represent a new family of ternary Mn-based superconductors with relatively high $T_c$. More studies are needed to reveal the complex magnetism and its relationship with the observed superconductivity.

## Methods

**Sample preparation and characterizations at AP.** RbMn$_6$Bi$_5$ single crystals were grown by using the Rb-Bi flux method as reported elsewhere[10]. We checked the single-crystal X-ray diffraction (XRD) at 300 K and confirmed that the RbMn$_6$Bi$_5$ crystallizes in a monoclinic structure with space group $C$2/m. The refined lattice parameters are basically consistent with the previous report. Energy dispersive X-ray spectroscopy (EDX) measurements on the fresh surface of crystals confirm that the average chemical composition is close to the stoichiometric one. Electrical transport and magnetic properties at ambient pressure were measured on the Quantum Design Physical Property Measurement System (PPMS) and Magnetic Property Measurement System (MPMS-III), respectively.

**High-pressure resistivity and AC magnetic susceptibility.** High-pressure transport and AC magnetic susceptibility were performed in the palm CAC apparatus with high hydrostatic pressures[19,20]. Standard four-probe method was employed for resistivity measurement with the current applied along the $b$-axis and the magnetic field perpendicular the $b$-axis. Glycerol was employed as the pressure transmitting medium (PTM). The pressure values inside the CAC were determined from the superconducting transition of Pb at low temperatures. AC magnetic susceptibility was measured by mutual induction method at a fixed frequency in a CAC apparatus. The primary and secondary coils are made of enameled cooper wires of 25 μm in diameter. All the low-temperature experiments were performed in a $^4$He refrigerated cryostat equipped with a 9 T superconducting magnet at the Synergic Extreme Condition User Facility (SECUF). BeCu-type diamond pressure cell (DAC) with 300-μm flat was used to measure high-pressure resistance with KBr as the solid PTM. The pressure in DAC was monitored at room temperature with the ruby fluorescence method.

## Data availability

The data that support the findings of this study are available from the corresponding authors upon reasonable request.

## Acknowledgements

This work is supported by the Beijing Natural Science Foundation (Z190008), National Key R&D Program of China (2018YFA0305700), the National Natural Science Foundation of China (Grant Nos. 12025408, 11874400, 11834016, 11921004, 11888101), the Strategic Priority Research Program and Key Research Program of Frontier Sciences of CAS (XDB25000000, XDB33000000 and QYZDB-SSW-SLH013), the CAS Interdisciplinary Innovation Team (JCTD-201-01) and the Users with Excellence Program of Hefei Science Center CAS (2021HSC-UE008). IOP Hundred-Talent Program (Y7K5031X61), and Youth Promotion Association of CAS (2018010). Y.U. acknowledges the support from JSPS KAKENHI (Grant No. JP19H00648).

apparatus. *Rev. Sci. Instrum.* **85**, 093907 (2014).

**Figure captions**

**FIG. 1** (a) Crystal structure of RbMn$_6$Bi$_5$. (b) The resistivity and magnetic susceptibility along the *b*-axis of RbMn$_6$Bi$_5$ at AP. The $T_N$ shows the antiferromagnetic transition temperature. The Curie-Weiss fitting is shown by the broken line.

**FIG. 2** (a) Electrical resistance of RbMn$_6$Bi$_5$ under various pressures measured in a CAC. (b) The low-temperature $R(T)$ data highlighting the superconducting transition. (c) The temperature derivative d$R$/d$T$ showing the evolutions of the antiferromagnetic transition.

**FIG. 3** (a) $R(T)$ at 14.5 GPa under different magnetic fields. (b) The temperature dependences of $\mu_0 H_{c2}(T)$ fitted by the Ginzburg-Landau (GL) equation, $\mu_0 H_{c2}(T) = \mu_0 H_{c2}(0)[1-(T/T_c)^2]/[1+(T/T_c)^2]$. (c) The ac magnetic susceptibility $\chi'(T)$ of RbMn$_6$Bi$_5$ (#2) and a piece of Pb with similar volume. (d) The low-temperature $R(T)$ data of RbMn$_6$Bi$_5$ (#3) measured with a DAC.

**FIG. 4** *T-P* phase diagram of RbMn$_6$Bi$_5$. The AFM and SC refers to the antiferromagnetic and the superconducting phases, respectively.

# Figure 1

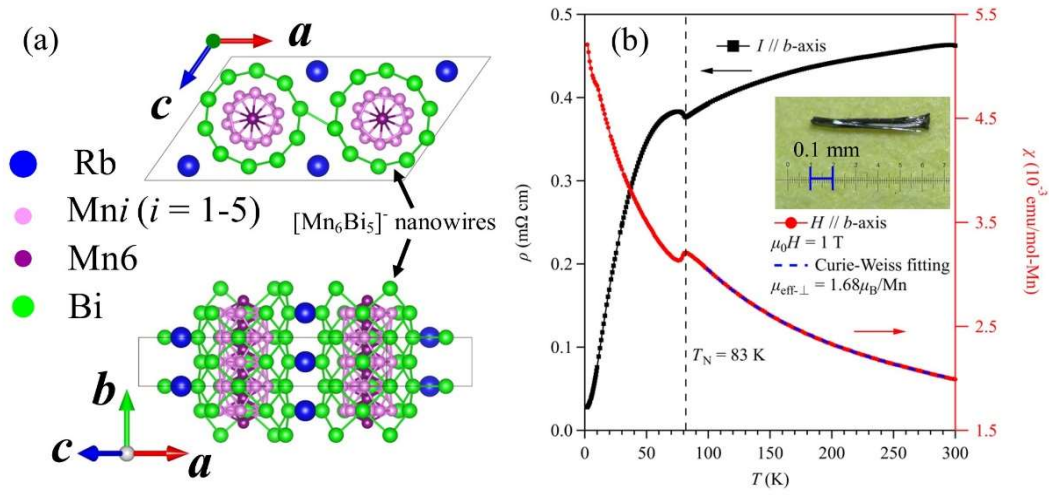

# Figure 2

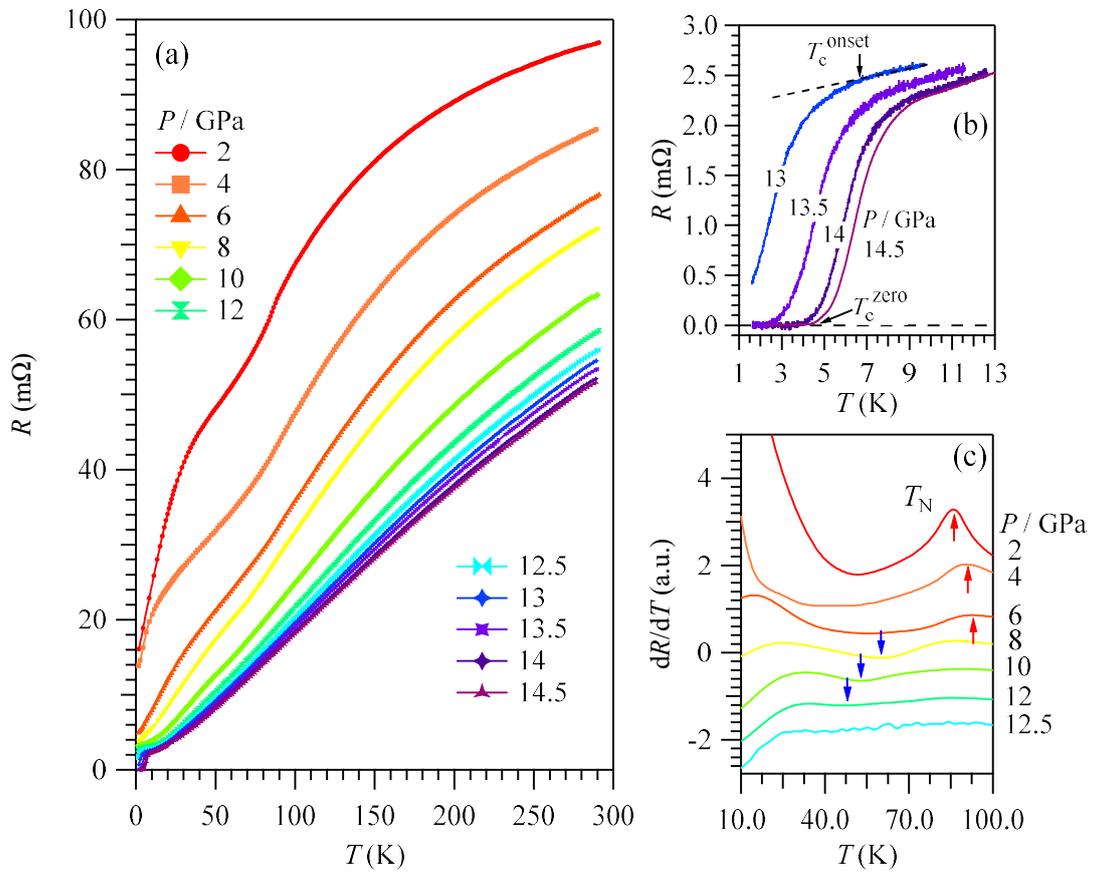

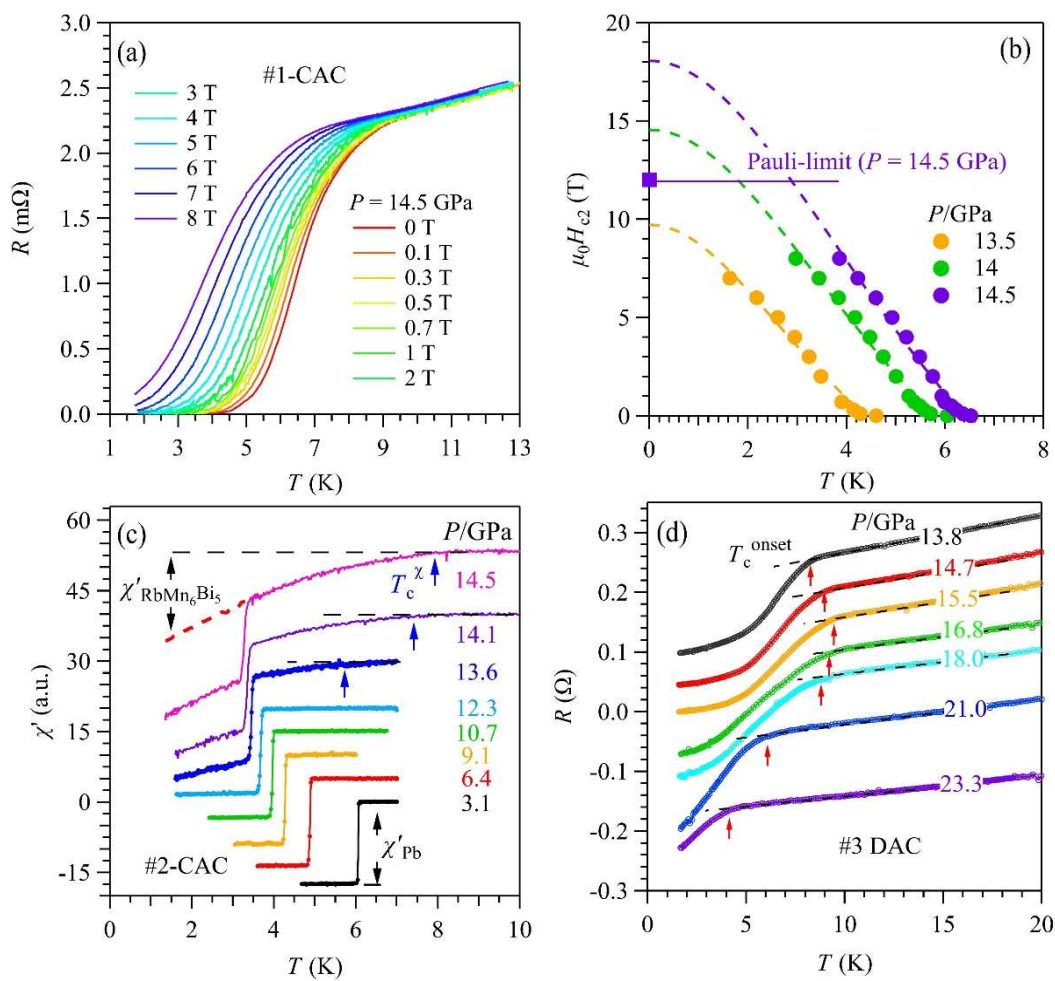

**Figure 4**

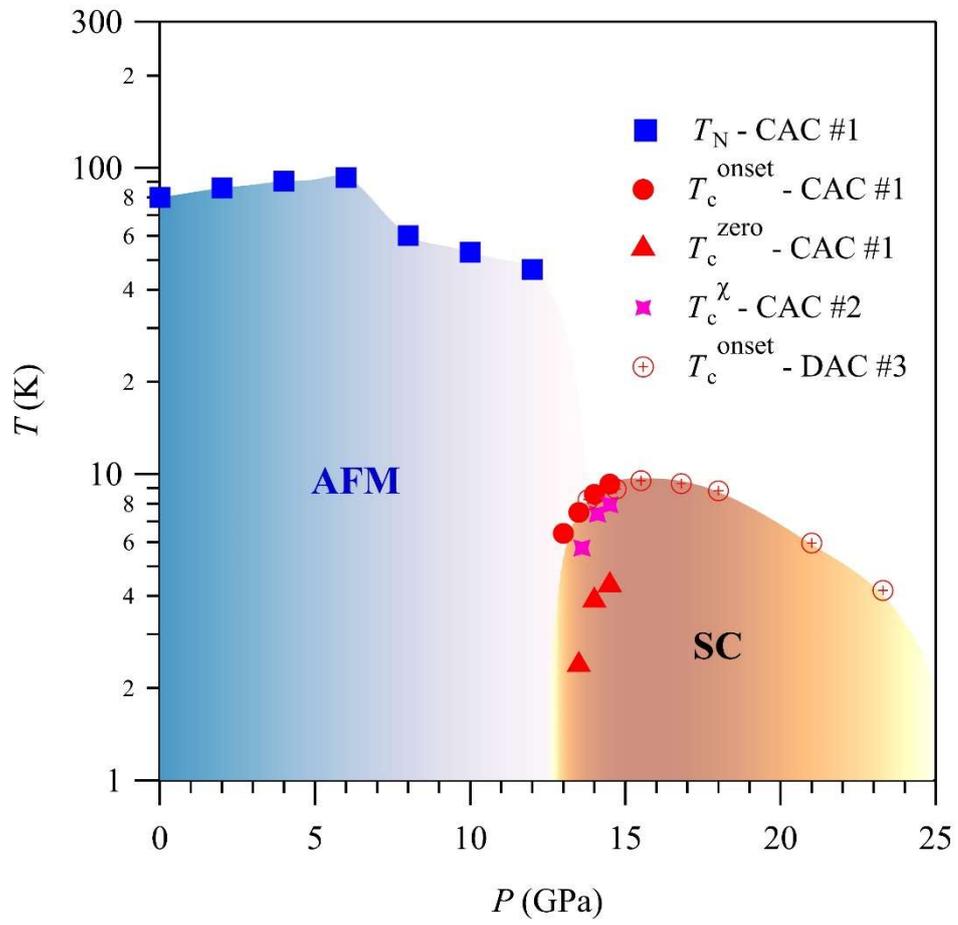

**Figure S1**

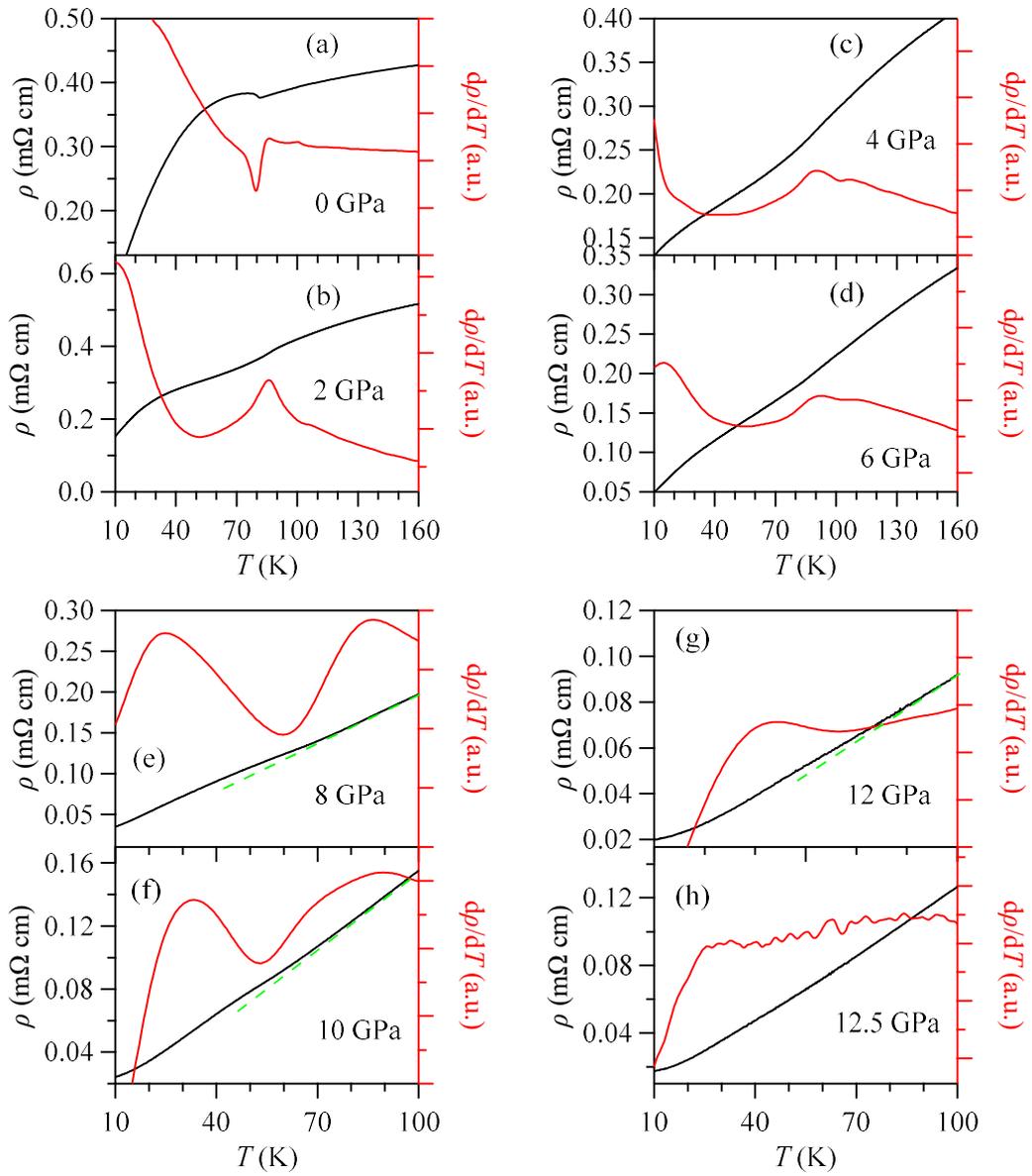

- 13 -

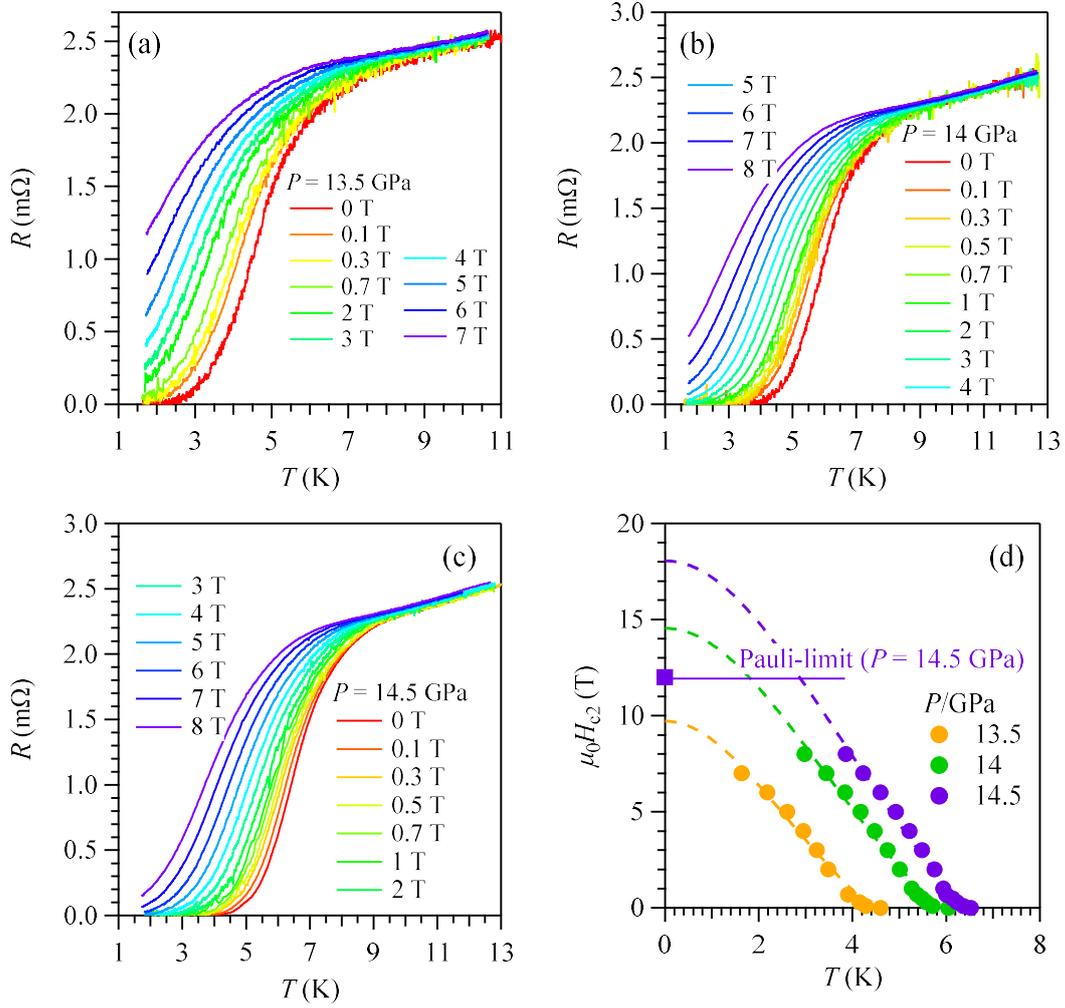